\documentclass[conference]{IEEEtran}
\IEEEoverridecommandlockouts

\usepackage{lineno,hyperref}

\usepackage[utf8]{inputenc} 
\usepackage[T1]{fontenc}
\usepackage{url}
\usepackage{ifthen}
\usepackage{amssymb}
\usepackage[cmex10]{amsmath} 
\usepackage{epsfig}
\usepackage{amsfonts}
\usepackage{pict2e}
\usepackage{color}
\usepackage{tikz}
\usepackage{pgfplots}

\def\qed{\hfill $\blacksquare$}

\newtheorem{Theorem}{Theorem}
\newtheorem{Definition}{Definition}
\newtheorem{Lemma}{Lemma}

\newtheorem{Remark}{Remark}

\def\M{{\mathcal M}}

\def\NN{{\mathbb N}}

\newcommand{\IEEEauthorblockLTI}[1]{%
	\IEEEauthorblockA{%
		\IEEEauthorrefmark{1}%
		\textit{Lehr- und Forschungseinheit f\"ur Nachrichtentechnik} \\
		\textit{ Technische Universit\"at M\"unchen}\\
			Munich, Germany \\ #1}
}
\newcommand{\IEEEauthorblockLNT}[1]{%
	\IEEEauthorblockA{%
		\textit{Kharkevich Institute for Information Transmission Problems} \\
		\textit{Russian Academy of Sciences}\\
		Moscow, Russia \\ #1}
}

\title{How to apply the rubber method for channels with feedback%
	\thanks{{\fontsize{5}{4}\selectfont Christian Deppe was supported by the Bundesministerium 
			f\"ur Bildung und Forschung (BMBF) through Grant 16KIS1005. Vladimir Lebedev was supported in part by the Russian Foundation for Basic Research, project no. 19-01-00364 and project no. 20-51-50007. Georg Maringer’s work was supported by the German Research Foundation (Deutsche Forschungsgemeinschaft, DFG) under Grant No.WA3907/4-1.}}}

\author{%
	\IEEEauthorblockN{Christian Deppe\IEEEauthorrefmark{1}, Georg Maringer\IEEEauthorrefmark{1}}
	\IEEEauthorblockLTI{christian.deppe@tum.de, georg.maringer@tum.de}
	\and
	\IEEEauthorblockN{Vladimir Lebedev}
	\IEEEauthorblockLNT{lebedev37@mail.ru}
}

\begin{document}
	\maketitle
	
	\begin{abstract}
	We give an overview of applications of the rubber method. The rubber method is a coding algorithm that was developed in 2005 by Ahlswede, Deppe and Lebedev for channels with feedback. It was a big surprise that an encoding strategy that reserves one symbol exclusively as a rubber can sometimes reach the capacity of a channel. Since then, generalizations of the algorithm have been developed. These generalizations can be used for special channels. There are also other ideas for modifying the rubber method.
	\end{abstract}
	

\section{Introduction}\label{introduction}

The rubber method is a coding algorithm for channels with feedback. It was introduced in 2005 in the work \cite{ADL05}. The authors discovered this algorithm when looking for a coding algorithm for a general $q$-ary error-correcting code with feedback that achieves the capacity error function. It was a surprise that this is possible for algorithms that reserve one symbol for error correction, meaning that the symbol cannot be used for encoding the information of the message. This is the main idea of the rubber method. Here, exactly one symbol is reserved as a rubber. Before the work \cite{ADL05} appeared, only the capacity error function for the case $q = 2$ was known. This problem was solved by Berlekamp \cite{B68} and Zigangirov \cite{Z76}.

A generalization of the rubber method was developed in \cite{L16} by Lebedev to get better rates for q-ary error-correcting codes. His idea was to encode information into the rubber while still using it to signal that an error occured.
In the meantime, the idea of the rubber method has also proven useful for other models. The work \cite{DLM20} shows that a generalized rubber method is also suitable for channels with unidirectional errors.
It can also be shown that the rubber method for the Z-channel with feedback can achieve a better rate compared to the same channel without feedback.
In this overview we want to concentrate on the rubber method, its generalizations and its applications. Our goal is to make more researchers aware of this simple and effective coding algorithm and to motivate them to work on further applications and generalizations.
In the first section we introduce the general model of error-correcting codes with feedback and the capacity error function for the binary case. In the second section we consider the original application of the rubber method for $q$-ary error-correcting codes with feedback. We define the rubber method and explain how to partially achieve the capacity error function. In the third section we explain the generalization of the rubber method of Lebedev \cite{L16}. In the fourth section we consider unidirectional errors and present the modification of the rubber method presented in \cite{DLM20}. Furthermore, we show that with this modification we also achieve a good algorithm for the Z-channel. We end this overview with a look at further applications of the rubber method and possible further extensions.

\section{Error-Correcting Codes with feedback}\label{feedback}

\begin{figure}[h]
	\centering
	\setlength{\unitlength}{0.72 cm}
	\begin{picture}(12,6)
	\put(1,1){\framebox(2.5,1){SENDER}}
	\put(5,1){\framebox(2.5,1){CHANNEL}}
	\put(9,1){\framebox(2.5,1){RECEIVER}}
	\put(3.5,1.5){\vector(1,0){1.5}}
	\put(7.5,1.5){\vector(1,0){1.5}}
	\put(8.25,1.5){\line(0,1){3.5}}
	\put(8.25,5){\vector(-1,0){4}}
	\put(2.25,5){\line(1,0){4}}
	\put(2.25,2){\line(0,1){3}}
	\put(4.5,4.5){feedback}
	\put(5,2.5){noise}
	\put(6,2.5){\line(0,-1){1.25}}
	\put(6,1.25){\line(1,1){0.5}}
	\put(6.5,1.75){\vector(0,-1){1.5}}
	\end{picture}
	\caption{Channel with feedback}
	\label{fig:channel_feedback}
\end{figure}
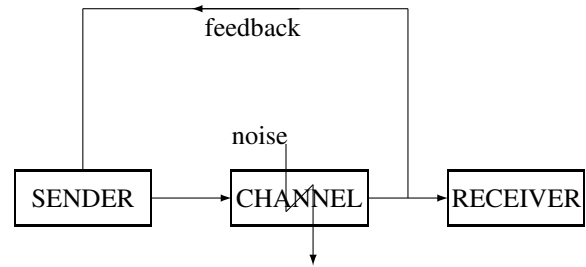

In this work we consider the following model of a transmission scheme.
A sender wants to send a message contained in the set $\M=\{1,2,\dots,M\}$ over a channel with noiseless feedback (see Figure \ref{fig:channel_feedback}). 
This channel has a $q$-ary alphabet $\mathcal{X} = \mathcal{Q} = \{0,\dots,q-1\}$ at its input and at its output $\mathcal{Y} = \mathcal{Q}$. Each message $m$ is encoded into a block of length $n$.
The channel is noisy and we assume that $t$ of the $n$ symbols can
be changed by the channel during the transmission.

\begin{Definition}\label{def:feedback_encoding}
\begin{enumerate}
	\item Let the set of possible messages be denoted as $\mathcal{M} = \{1,\dots,M\}$. Then an encoding algorithm for a feedback channel with blocklength $n$ is a set of functions
	
	\begin{equation}
	c_i: \mathcal{M} \times \mathcal{Q}^{i-1} \rightarrow \mathcal{Q}, \quad i \in \{1,\dots,n\} \enspace .
	\end{equation}
	
	The respective encoding algorithm is then constructed as
	\begin{equation}
	c(m,y^{n-1}) = ((c_1(m), c_2(m,y_1), \dots, c_n(m,y^{n-1})) \enspace ,
	\end{equation}
	where $y^k = (y_1,\dots,y_k)$.
	\item An encoding algorithm is $(n,M,t)_f$-successful if the corresponding decoder decodes correctly for any transmitted message $m$ and any error pattern with less than or equal $t$ errors caused by the channel.
	\end{enumerate}
\end{Definition}

Definition \ref{def:feedback_encoding} shows that the encoder may adjust its algorithm for sending the $k$th symbol $c_k$ according to the previously received symbols $y^{k-1}$. This extra flexibility can be used to increase the achievable rate of the system.
The rate of a  $(n,M,t)_f$ encoding algorithm is defined as $R=\frac {\log_q M}n$.
\begin{Definition}
	Let $2\leq q\in\NN$. The capacity error function $C_q^f(\tau)$ of a channel with noiseless feedback is defined as the supremum on the rates for which a successful algorithm exists depending on $\tau=\frac{t}{n}$ as the blocklength $n$ goes to infinity. $t$ denotes the maximum number of errors inflicted by the channel noise and $q$ specifies the alphabet size at the input and the output of the channel.

\end{Definition}
If the sender and the receiver share a general channel with noiseless passive feedback (Figure \ref{fig:channel_feedback}), the capacity error function is only completely known for the binary case.

Berlekamp first considered this problem in \cite{B68}. He showed the following upper bound for the error-correcting capacity function.

\begin{figure}[h]\tiny
    \centering
   \setlength{\unitlength}{0.5 cm}
\begin{picture}(12.5,8)
\put(0.6,7.4){$C_2^f({\tau})$}
\put(12.2,0.9){$\tau$}
\put(0.9,1){\vector(1,0){11.1}}
\put(1,0.9){\vector(0,1){6.1}}
\put(0.5,0.5){0}
\put(3.05,0.9){\line(0,1){0.2}}
\put(2.75,0.5){0.1}
\put(5.1,0.9){\line(0,1){0.2}}
\put(5,0.9){\line(0,1){0.2}}
\put(4.8,0.5){$\tau_t$}
\put(7.15,0.9){\line(0,1){0.2}}
\put(6.85,0.5){0.3}
\put(8,0.9){\line(0,1){0.2}}
\put(7.8,0.5){$\frac 13$}
\put(1,1){\dashbox{0.1}(4,1.5)}
\put(9.2,0.9){\line(0,1){0.2}}
\put(8.9,0.5){0.4}
\put(11.25,0.9){\line(0,1){0.2}}
\put(10.95,0.5){0.5}
\put(0.9,2){\line(1,0){0.2}}
\put(0.2,1.9){0.2}
\put(0.9,3){\line(1,0){0.2}}
\put(0.2,2.9){0.4}
\put(0.9,4){\line(1,0){0.2}}
\put(0.2,3.9){0.6}
\put(0.9,2.5){\line(1,0){0.2}}
\put(0.2,2.4){$R_t$}
\put(0.9,4.5){\line(1,0){0.2}}
\put(0.2,4.4){$R_0$}
\put(0.9,5){\line(1,0){0.2}}
\put(0.2,4.9){0.8}
\put(0.9,6){\line(1,0){0.2}}
\put(0.5,5.9){1}
\put(8,1){\line(-2,1){3}}
\bezier{500}(1,6)(1.5,4)(5,2.5)
\end{picture}
    \caption{Capacity error function for binary error correcting codes with feedback}
    \label{binary}
\end{figure}
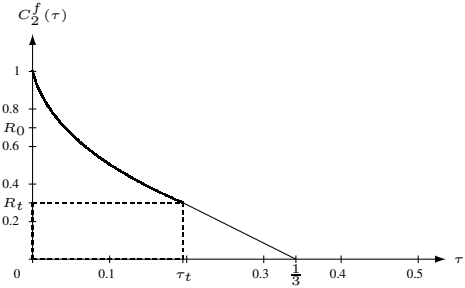

\begin{Theorem}[Upper Berlekamp Bound]
     Let $\tau, C_2^f(\tau)$ be defined as before, then it holds:
$$ C_2^f \leq
\begin{cases} 
1-h(\tau) &, \text{if } 0\leq \tau\leq \tau_c \\
(1- 3\tau)R_0 &, \text{if } \tau_c\leq \tau\leq {\frac 13},\end{cases}
$$
where $R_0=\log_2 \frac {1+\sqrt 5}2$ and $\tau_c=(3+\sqrt 5)^{-1}$
\end{Theorem}
Furthermore, Berlekamp showed the following theorem:
\begin{Theorem}[Lower Bound, Berlekamp]\label{eq:symmetric_channel}
     Let $\tau, C_2^f(\tau)$ be defined as before, then it holds:
$$ C_2^f \geq (1- 3\tau)R_0,$$
if $\tau_c\leq \tau\leq {\frac 13}$,
where $R_0=\log_2 \frac {1+\sqrt 5}2$ and $\tau_c=(3+\sqrt 5)^{-1}.$
\end{Theorem}
Later it turned out that a much easier proof of this achievability 
can be done by the rubber method. We show this in the next section.
In \cite{Z76} Zigangirov proved the following result:
\begin{Theorem}[Lower Zigangirov Bound]
     Let $\tau, C_2^f(\tau)$ be defined as before, then it holds:
$$ C_2^f(\tau) \geq 1-h(\tau)$$
if $0\leq \tau\leq \tau_c$, where $\tau_c=(3+\sqrt 5)^{-1}$.
\end{Theorem}
Therefore Berlekamp's upper bound is achievable. The capacity error function $C_2^f(\tau)$ is shown in Figure~\ref{binary}. In the
next section we will show how we can partly reach this capacity error function
with the rubber method.

\section{Non-binary error-correcting codes with feedback and the rubber method}

In the previous section we stated that the capacity error function with feedback is known in the binary case. In their paper \cite{ADL05}, Ahlswede, Deppe and Lebedev then dealt with the question of finding the capacity error function for $q>3$. With the generalization of Berlekamp's translation bound (see also Aigner \cite{A95}) and the q-ary Hamming bound, they obtained the following upper bound.
\begin{Theorem}\label{ADLu}
    Let $\tau, C_q^f(\tau)$ be like before, then holds:
$$ C_q^f(\tau) \leq
\begin{cases} 
1- h_q(\tau ) - \tau\log_q(q-1) &, \text{if } 0\leq \tau\leq \frac 1q \\
(1-2\tau)\log_q(q-1) &, \text{if } \frac 1q\leq \tau\leq {\frac 12},\end{cases}
$$
where $h_q(\tau)=-\tau\log_q(\tau)-(1-\tau)\log_q(1-\tau)$.
    
\end{Theorem}
To show the achievability of the bound, Ahlswede, Deppe and Lebedev developed the rubber method in \cite{ADL05}. This coding algorithm achieves the upper bound in Theorem~\ref{ADLu} 
for relative errors ${\frac 1q}\leq \tau \leq {\frac 12}$.
 
It was shown that it is possible to transmit $(q-1)^{n-2t}$ messages in a block length $n$. A bijection $b$
of
messages $\M$ to the set $\{1,2,\dots, q-1\}^{n-2t}$ of used sequences
is agreed upon by the sender and the receiver.

Given message $i\in\M$, the sender chooses $b(i)=x^{n-2t}=(x_1,x_2,\dots,x_{n-2t})\in \{1,2,\dots, q-1\}^{n-2t}$
as a {\bf skeleton for encoding}. The receiver's goal is to get
the skeleton and therefore the message.
The ``0'' is used for error correction only. If the message has been correctly received and there are still symbols remaining within the block, then those are filled by ``1`` symbols from the transmitter. 

{\bf Transmission algorithm:}
 
The sender sends $x_1$, continues with $x_2$ and so on until the first error occurs, say in position $p$ with $x_p$ sent.
The error here can be of two kinds: a {\bf standard error} (that means symbol $x_p$ is changed to another symbol $y_p\in\{1,2,\dots, q-1\}$)
and a {\bf towards zero error} (that means $x_p$ is changed to $y_p=0$).

If a {\bf standard error} occurs, the sender transmits with smallest $l$ possible, $2l+1$ times $0$
(where $l\in\NN\cup 0$) until the decoder receives
$l+1$ zeros (known to the sender via feedback. Such an $l$ exists because the number of errors is bounded by $t$). 
Then at the next
step he transmits 
$x_p$ again, and continues the algorithm.\\
 
 
 
If a {\bf towards zero error} occurs, the sender decreases $p$ by one
(if it is bigger than 1) and continues (transmits $x_p$ at the next step).
 

{\bf Decoding algorithm:}
 
 
The decoding is very simple. The receiver just regards the ``0'' as a
kind of deletion
symbol - he erases it by a rubber, {\bf which in addition erases the previous symbol.}\\
 
This is the reason why the sender has to repeat sending the symbol according to the skeleton if a
towards zero error occurs. 

At the end the first $n-2t$ symbols at the decoder are those of $b(i)=(x_1,x_2,\dots,x_{n-2t})$.

Indeed, suppose that $t_0$ towards zero errors occur. The necessary retransmissions and the unintentional zero symbols reduce the effective block length by $2t_0$ symbols.

So we are left with $t_1=t-t_0$ possible errors and a block length $n-2t_0$, and standard errors as well as
correction errors, meaning a change of an (intentional) zero symbol to a non-zero symbol. The standard errors $s_1,\dots,s_r$ cause
correction errors $l_1,\dots, l_r$, respectively, and a loss in block length of  $2(l_1+1),\dots, 2(l_r+1)$, respectively, and thus in total 
$\sum_{i=1}^r (1+l_i)\leq t_1$ errors and a reduction of $\sum_{i=1}^r 2(l_i+1)\leq 2t$ symbols of the block length.

Hence within the block $n-2(t_0+t_1)=n-2t$ symbols are guaranteed to be successfully transmitted over the channel with our strategy, corresponding to $M=(q-1)^{n-2t}$ messages.

Thus ${\log_q \frac Mn}=(1-{\frac {2t}n})\log_q (q-1)$ and we have derived the result of 
Ahlswede, Deppe and Lebedev.
\begin{Theorem}
For $\tau={\frac tn}$ and $0<\tau<{\frac 12}$
$$C^f_q(\tau)\geq (1-2\tau)\log_q(q-1).$$
\end{Theorem}

\bigskip
Now we will show that we can generalize the rubber method in such a way that we get as the rate function a tangent
to the Hamming bound through $(\tau,0)=({\frac 1{r+1}},0)$, where $1\leq r\in\NN$. 
The generalization also works in this case for $q=2$ and gives an alternative optimal encoding strategy for Berlekamp's method.
The $r$-rubber method is defined as follows: The communicators map all messages to sequences of the set
${\cal X}_r^{n-(r+1)t}=\{ x^{n-(r+1)t} : \mbox{the sequence contains $\leq r-1$ consecutive zeros}\}$
and the sender uses $r$ consecutive zeros as a deletion sequence.
The following theorem is shown in \cite{ADL05}.
\begin{Theorem}

The rate function of the $r$-rubber method is a tangent to $H_q(\tau)$ going through ${1\over r+1}$.
\end{Theorem}
\section{Lebedev's generalized rubber method}

Consider the rubber method for $r= 1$. Previously, an information sequence $(m_1,m_2,\dots,m_{n-2t})$ did not contain zeros, and $2t$ positions were used for error correction. Lebedev \cite{L16} considered an information sequence of the form
$m=(m_1,m_2,\ldots ,m_{n-2t-\lceil \log_{q-1}\binom{z+t}{z}
\rceil})$, containing precisely $z$ zeros. Within his algorithm he additionally uses a check sequence consisting only of nonzero elements. We use this sequence to enumerate $\binom{z+t}{z}$ variants for the location of 
$z$ information zeros among, at most, $z+t$ zeros. For that we need $\lceil
\log_{q-1}\binom{z+t}{z} \rceil $ information symbols.
The check sequence is in the following denoted as $h=(h_1,h_2,\ldots ,h_{\lceil
\log_{q-1}\binom{z+t}{z} \rceil})$. 
Thus, an information sequence will consist of $n-2t-\lceil \log_{q-1}\binom{z+t}{z} \rceil $ symbols and a check sequence of $\lceil \log_{q-1}\binom{z+t}{z} \rceil $ symbols.

The encoder will transmit the information sequence using the above-described 1-rubber algorithm with the only difference being 
that it will keep track of which zero symbols belong to the information symbols, the intentionally sent rubber symbols, as well as zero symbols that have been created by the channel without the sender's intention. Information zeros do not function as rubber symbols and therefore those and the symbols before them are not erased. There is, of course, no retransmission necessary in that case. We will formalise this coding strategy in the following:

The $1$-rubber method (with the modifications described in the previous paragraph) is used to transmit the information sequence $m$. During this procedure the sender counts the amount of zero symbols within the output sequence that are created by the channel without the sender's intention. We denote this number by $V(i)$, where $i$ labels how many symbols have already been transmitted.
Note that $V(i)\leq V(i+1)$ and that $V(i)$ may take different values at different times $i$.
After the message $m$ has been transmitted, say at time $j$, the sender transmits a sequence of $V(j)$ ones using the $1$-rubber method. If an error leading the zero symbol within the output sequence occurs, an additional $1$ symbol has to be sent.

Finally the check sequence $h$ is transmitted using the $1$-rubber method. If there are still symbols remaining in the block, then those are filled up using $1$ symbols.

The entire sequence to be transmitted therefore looks like the following:

\begin{eqnarray*}
a^{\ast}&=&(m_1,m_2,\ldots ,m_{n-2t-\lceil \log_{q-1}\binom{z+t}{z}
\rceil},1,\ldots,1,\\
&&h_1,h_2,\ldots ,h_{\lceil
\log_{q-1}\binom{z+t}{z} \rceil},1,1,\ldots).
\end{eqnarray*}

The following theorem gives the maximal size of a message set than can be successfully transmitted over the channel by Lebedev's generalized rubber method:

\begin{Theorem} The generalized 1-rubber algorithm transmits
$$ \binom{n-2t-\lceil \log_{q-1}\binom{z+t}{z} \rceil}{z}\cdot (q-1)^{n-2t-\lceil \log_{q-1}\binom{z+t}{z} \rceil -z}$$
messages.
\end{Theorem}

Of course the problem of how much the new algorithm improves the previous one is of interest. This problem is not so easy and requires further investigation. In \cite{L16}, Lebedev gave some numerical results which show that the algorithm does improve. However it did not reach the upper bound. The situation
becomes better if $q$ increases.

\section{Special channels and the rubber method}

In the previous sections we consider channels where an error
can change a symbol to every possible symbol. For some applications
one can assume that a symbol can, by an error, be changed only to a subset 
of the symbols. We therefore define a general discrete channel.
The general discrete channels can be specified by bipartite graphs.

\begin{Definition}[Discrete channel/bipartite graph]
	A discrete channel corresponds to a bipartite graph in the following way. Let $\mathcal{V}_{in}$ denote the set of possible input symbols and let $\mathcal{V}_{out}$ denote the set of possible output symbols. Then if it is possible that the channel maps input symbol $i \in \mathcal{V}_{in}$ to output symbol $j \in \mathcal{V}_{out}$, the pair $(i,j)$ is part of the set of edges \\ $\mathcal{E} \subset \mathcal{V}_{in} \times \mathcal{V}_{out}$. Conversely, a bipartite graph between input symbols $\mathcal{V}_{in}$ and $\mathcal{V}_{out}$ defines a discrete channel.
	
		We denote by $(x_1,x_2,\dots,x_n)$ the sequence of input symbols and by $(y_1,y_2,\dots, y_n)$ the sequence of output symbols, where $x_i$ depends
	on the message and $(y_1,\dots,y^{i-1})$ (see Definition~\ref{def:feedback_encoding}).
	We define $(e_1,e_2,\dots, e_n)$ as the corresponding error vector, where $e_i=y_i-x_i$.
\end{Definition}

Discrete channels with finite input and output alphabets correspond to their respective bipartite graphs by a one to one mapping. Therefore, in this paper we frequently speak about the graphs when we mean the respective channels and vice versa.

\begin{Definition}[Asymmetric channel]
	An asymmetric channel is a discrete channel whose specifying bipartite graph is not the full graph in the sense that the set of edges $\mathcal{E} \neq \mathcal{V}_{in} \times \mathcal{V}_{out}$.
\end{Definition}

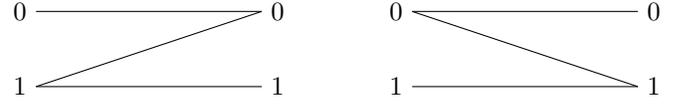
\begin{figure}
	\centering
	\begin{tikzpicture}
	\draw (0,0) node [left] {$0$} -- (3,0) node [right] {$0$};
	\draw (0,-1) -- (3,0);
	\draw (0,-1) node [left] {$1$} -- (3,-1) node [right] {$1$};
	\draw (5,0) node [left] {$0$} -- (8,0) node [right] {$0$};
	\draw (5,0) -- (8,-1);
	\draw (5,-1) node [left] {$1$} -- (8,-1) node [right] {$1$};
	\end{tikzpicture}
	\caption{binary Z-channel and inverse Z-channel}
	\label{fig:binary_Z_channel}
\end{figure}

As an example of an asymmetric channel we propose the Z-channel, which is specified by the bipartite graph on the left hand side of Figure \ref{fig:binary_Z_channel}.


\begin{Definition}[Unidirectional channel]
	A unidirectional channel is specified by two asymmetric channels having the same set of input symbols $\mathcal{V}_{in}$ and output symbols $\mathcal{V}_{out}$. One of the channels allows only positive error vectors $(e_i \geq 0 \; \forall i)$, whereas the other one only allows negative error vectors $(e_i \leq 0 \; \forall i)$. Within each transmission of a codeword, the channel is specified by one of the asymmetric channels. Sender and receiver know both channels, but they do not know to which asymmetric channel the unidirectional channel corresponds. This may change for each codeword.
\end{Definition}

In Figure \ref{fig:binary_Z_channel} the binary unidirectional channel is shown. It is denoted as $\Gamma_U^2$ and is composed of the binary Z-channel and its counterpart, the inverse Z-channel.


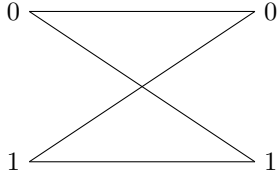
\begin{figure}
	\centering
	\begin{tikzpicture}
	\draw (0,0) node [left] {$0$} -- (3,0) node [right] {$0$};
	\draw (0,-2) -- (3,0);
	\draw (0,0) -- (3,-2);
	\draw (0,-2) node [left] {$1$} -- (3,-2) node [right] {$1$};
	\end{tikzpicture}
	\caption{Symmetric channel $\Gamma^2$}
	\label{fig:symmetric_channel}
\end{figure}

\begin{Definition}[Generalized Z/inverse Z-channel]
	The generalized Z-channel $\Gamma_Z^q$ is specified by the bipartite graph with input nodes $\mathcal{V}_{in} = \mathcal{V}_{out} = \{0,\dots, q-1\}$ and the set of edges $\mathcal{E}^q_Z = \{(i,i-1) : i \in \{1,\dots,q-1\}\} \cup \{(i,i) : i \in \{0,\dots,q-1  \}\}$.
	
	The inverse Z-channel $\Gamma_{\reflectbox{$\mathsf{Z}$}}^q$ is specified by the bipartite graph with input nodes $\mathcal{V}_{in} = \mathcal{V}_{out} = \{0,\dots, q-1\}$ and the set of edges $\mathcal{E}^q_{\reflectbox{$\mathsf{Z}$}} = \{(i,i+1) : i \in \{0,\dots,q-2\} \cup \{(i,i) : i \in \{0,\dots,q-1 \}  \}$.
\end{Definition}

The probabilistic model of the $q$-ary case was considered in \cite{D75} and an upper bound on the error probability was given.

\begin{Remark}\label{rem:Z_inverse_Z}
	The capacity error function of the generalized Z-channel $\Gamma_Z^q$ is equal to the capacity error function of the inverse generalized Z-channel $\Gamma_{\reflectbox{$\mathsf{Z}$}}^q$.
\end{Remark}

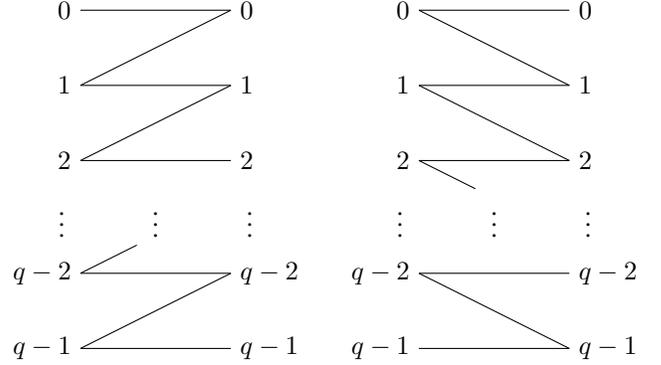
\begin{figure}
	\centering
	\begin{tikzpicture}
	\draw (0,0) node [left] {$0$} -- (2,0) node [right] {$0$};
	\draw (0,-1) -- (2,0);
	\draw (0,-1) node [left] {$1$} -- (2,-1) node [right] {$1$};
	\draw (0,-2) -- (2,-1);
	\draw (0,-2) node [left] {$2$} -- (2,-2) node [right] {$2$};
	\draw (0,-3.5) -- (0.75, -3.125);
	\node at (1, -2.75) {$\vdots$};
	\node at (-0.25, -2.75) {$\vdots$};
	\node at (2.25,-2.75) {$\vdots$};
	\draw (0,-3.5) node [left] {$q-2$} -- (2,-3.5) node [right] {$q-2$};
	\draw (0,-4.5) -- (2,-3.5);
	\draw (0,-4.5) node [left] {$q-1$} -- (2,-4.5) node [right] {$q-1$};
	
	\draw (4.5,0) node [left] {$0$} -- (6.5,0) node [right] {$0$};
	\draw (4.5,0) -- (6.5,-1);
	\draw (4.5,-1) node [left] {$1$} -- (6.5,-1) node [right] {$1$};
	\draw (4.5,-1) -- (6.5,-2);
	\draw (4.5,-2) node [left] {$2$} -- (6.5,-2) node [right] {$2$};
	\draw (4.5,-2) -- (5.25, -2.375);
	\node at (4.25, -2.75) {$\vdots$};
	\node at (5.5, -2.75) {$\vdots$};
	\node at (6.75,-2.75) {$\vdots$};
	\draw (4.5,-3.5) node [left] {$q-2$} -- (6.5,-3.5) node [right] {$q-2$};
	\draw (4.5,-3.5) -- (6.5,-4.5);
	\draw (4.5,-4.5) node [left] {$q-1$} -- (6.5,-4.5) node [right] {$q-1$};
	\end{tikzpicture}
	\caption{Generalized Z-channel and generalized inverse Z-channel}
	\label{fig:Z_channel}
\end{figure}

Both channels are depicted in Figure \ref{fig:Z_channel}. Combined they form a unidirectional channel which we denote as $\Gamma_{U}^q$.

The symbols $0$ and $q-1$ are of special interest for the generalized Z-channel and the inverse generalized Z-channel. For the generalized Z-channel, the symbol $0$ has the property that the sender knows that this symbol cannot be changed by the channel. If the symbol $q-1$ is received, the receiver knows that the transmitter indeed sent this symbol. Therefore we use a sequence of $r$ symbols with the value $q-1$ as the rubber. The properties of $0$ and $q-1$ are swapped for the inverse generalized Z-channel compared to the generalized Z-channel. All other symbols do not have these special properties.

Now we want to show that we can construct lower bounds using 
a modified rubber method.
In the previous section we showed that for the rubber method we could have two kinds of errors:
a {\bf standard error} (which means a symbol is changed to another symbol and
the sender sends the rubber sequence) and a {\bf towards rubber error} (which means a symbol is changed to a rubber symbol such that the receiver obtains a rubber sequence). If a {\bf towards rubber error} occurs, a correctly received symbol is deleted and has to be sent again.
For the generalized Z-channel $\Gamma_Z^q$, a {\bf towards rubber error}
is not possible if we use $r$ times $q-1$ as the rubber sequence. Also, for the
inverse generalized Z-channel $\Gamma_Z^q$, a {\bf towards rubber error}
is not possible if we use $r$ times $0$ as the rubber sequence. Therefore, the sender does not have to retransmit the previously erroneous symbol again, because for those channels there is only the possibility of a {\bf standard error}. We denote this modified algorithm (without retransmissions) by $A(r,b)$ if we use
$b^r=(b,\dots, b)$ as the rubber. 
\begin{Lemma}\label{th:rubber_retransmission}
	The modified rubber strategy $A(r,q-1)$ [$A(r,0)$] is a successful algorithm
	for the the generalized [inverse] Z-channel $\Gamma_Z^q$ [$\Gamma_{\reflectbox{$\mathsf{Z}$}}^q$].
\end{Lemma}
In the following, we denote this rubber method without retransmission as the modified rubber method.
To calculate the rate of this algorithm we need, as in \cite{ADL05}, the following definitions. Let $z_r^{r+1}=qz^r_r-q+ 1$. It is well known that for $n \rightarrow \infty $ the number of sequences of length $n$ not containing a block $b^r=(b,\dots, b)$ is asymptotically equal to $z_r^{n}$ (in
\cite{ADL05} and \cite{B77} we see how to choose the initial value for the iteration).

\begin{Theorem}\label{cor:rubber}
	Let $z_r$ be defined as above. Then for the generalized [inverse] Z-channel $\Gamma_Z^q$ [$\Gamma_{\reflectbox{$\mathsf{Z}$}}^q$] for $q \geq 2$ we get 
	 \begin{equation*}
        C_q^f(\Gamma_Z^q,\tau) \geq R_{mr}:= \begin{cases}
	\max\limits_{2\leq r\in\NN} (1-r\tau) \log_q z_r& \text{if } 0 \leq \tau \leq \frac{1}{2}\\
	0 & \text{if } \tau > \frac{1}{2}\, .
	\end{cases}
	\end{equation*}
\end{Theorem}
{\it Idea of the proof} : For the modified $r$-rubber method
we have
\begin{equation*}
	C_q^f(\Gamma_Z^q,\tau ) \geq (1-r\tau) \log_q z_r. 
	\end{equation*}
We use the modified rubber method achieving the highest asymptotic rate depending on the rubber length $r$ for each value of $\tau$. With this we obtain the lower bound $R_{mr}$.\qed


%

By a modification of the strategy used in Theorem~\ref{cor:rubber} we obtain the following result:

\begin{Theorem}\label{th:rubber_unidirectional}
	Let $\Gamma_U^q$ be a unidirectional channel consisting of the generalized Z-channel $\Gamma_Z^q$ and the inverse generalized Z-channel $\Gamma_{\reflectbox{$\mathsf{Z}$}}^q$ ($q \geq 2$). Then
	we have 
	\[
	C_q^f(\Gamma_U^q,\tau)\geq R_{mr}.
	\]
\end{Theorem}

\begin{Remark}
The values for $z_r$ in Theorem~\ref{cor:rubber} and Theorem~\ref{th:rubber_unidirectional} 
can be computed. An important case is $q=2$.
\end{Remark}

\begin{figure}
    \centering
    \scalebox{0.65}{
    \input{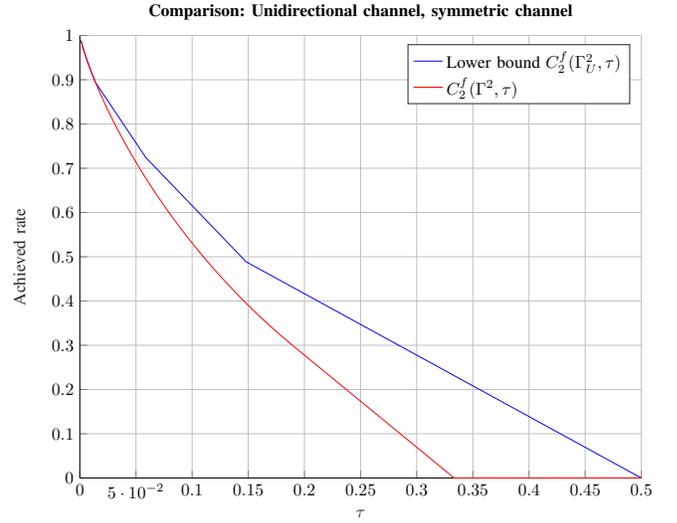}}
    \caption{Comparison of the lower bound on $C_2^f(\Gamma_U^2,\tau)$ obtained by the modified rubber method using the capacity error function of the symmetric channel $C_2^f(\Gamma^2,\tau)$}
    \label{fig:comparison_Berlekamp}
    \vspace{-1em}
\end{figure}

If we apply the modified rubber method to the binary case we get the following result.
The capacity error function of the unidirectional channel is lower bounded by
    \begin{equation*}
        C_2^f(\Gamma_U^2,\tau) \geq \begin{cases}
	\max\limits_{2\leq r\in\NN} (1-r\tau) \log_q z_r& \text{if } 0 \leq \tau \leq \frac{1}{2}\\
	0 & \text{if } \tau > \frac{1}{2}\, .
	\end{cases}
    \end{equation*}
The result for $r=2$ is
\[
    (1-2\tau)\log \left(\frac{1+\sqrt{5}}{2} \right).
    \]
The major change compared to the result given in Theorem~\ref{eq:symmetric_channel} is the change of the factor $(1-3\tau)$ to $(1-2\tau)$. It occurs because retransmissions after erroneous symbols are unnecessary.
Figure~\ref{fig:comparison_Berlekamp} shows the lower bound on $C_2^f(\Gamma_U^2,\tau)$ obtained by using the modified rubber method and $C_2^f(\Gamma^2,\tau)$ for comparison. This result is different without feedback, since the capacity error functions of the symmetrical and unidirectional channel are the same in that case.

\section{Conclusions}
This survey paper introduces and shows applications of the rubber method. The results show that although the technique itself is rather simple, the results obtained by it frequently achieve the capacity error function for certain channels. In other cases it can at least be used in a first step to obtain a reasonably good lower bound on the capacity error function of the channel. In this work the binary symmetric channel, some asymmetic channels as well as unidirectional channels have been considered, but the methodology may also be considered for other channels.

In the last section we considered a channel model
where the knowledge of an erroneous position at the receiver implies knowing the value of the respective symbol. Retransmissions are not necessary if the encoding strategy is chosen in a way such that the receiver is able to obtain the error positions. This can be used to create encoding strategies achieving a higher rate for many channels using the modified rubber method. Furthermore it was shown how to change the modified rubber method for unidirectional channels. The method proposed shows that it is possible to achieve the same rate for the unidirectional channel, consisting of a generalized Z-channel and an inverse generalized Z-channel as its components.
 It would be also interesting to analyze the rubber method for probabilistic channels.

\end{document}